\documentclass[a4paper]{article}
\usepackage{amscd,amsmath,amssymb}
\usepackage[usenames, dvipsnames]{color}

\newcommand{\R}{\mathbb R}

\newcommand{\p}[1]{(\ref{#1})}

\newcommand{\tF}{F^{(0)}}
\newcommand{\hF}{{\hat F}}
\newcommand{\htF}{{\hat F}^{(0)}}

\newcommand{\bQ}{{\overline Q}{}}

\newcommand{\bpsi}{{\bar\psi}{}}

\newcommand{\sfrac}[2]{{\textstyle\frac{#1}{#2}}}
\newcommand{\und}{\qquad\textrm{and}\qquad}
\renewcommand{\=}{\ =\ }

\newcommand{\be}{\begin{equation}}
\newcommand{\ee}{\end{equation}}
\newcommand{\bea}{\begin{eqnarray}}
\newcommand{\eea}{\end{eqnarray}}
\newcommand{\ba}{\begin{array}} \newcommand{\ea}{\end{array}}

\def\im{{\rm i}}
\def\pa{\partial}
\def\diff{{\rm d}}

\begin{document}

\pagenumbering{gobble}

\phantom{.}
\vspace{2cm}

\begin{center}
{\huge\bf Curved Witten--Dijkgraaf--Verlinde--Verlinde \\[10pt]
equation and ${\cal N}${=}\,4 mechanics}
\end{center}

\vspace{1cm}

\begin{center}
{\Large\bf  
Nikolay Kozyrev${}^{a}$, 
Sergey Krivonos${}^{a}$, 
Olaf Lechtenfeld${}^{b}$, \\[8pt]
Armen Nersessian${}^{c}$ and 
Anton Sutulin${}^{a}$
}
\end{center}

\vspace{1cm}

\begin{center}
${}^a$ {\it
Bogoliubov  Laboratory of Theoretical Physics, JINR,
141980 Dubna, Russia}

\vspace{0.2cm}

${}^b$ {\it
Institut f\"ur Theoretische Physik and Riemann Center for Geometry and Physics \\
Leibniz Universit\"at Hannover,
Appelstrasse 2, 30167 Hannover, Germany}

\vspace{0.2cm}

${}^c$ {\it Yerevan State University, 1 Alex Manoogian St., Yerevan, 0025, Armenia}

\vspace{0.6cm}

{\tt nkozyrev@theor.jinr.ru, krivonos@theor.jinr.ru, lechtenf@itp.uni-hannover.de,\\[4pt]
arnerses@ysu.am, sutulin@theor.jinr.ru}

\end{center}

\vspace{2cm}

\begin{abstract}\noindent
We propose a generalization of the Witten--Dijkgraaf--Verlinde--Verlinde (WDVV) equation from $\R^n$
to an arbitrary Riemannian manifold. Its form is obtained by extending the relation of the WDVV equation 
with ${\cal N}{=}\,4$ supersymmetric $n$-dimensional mechanics from flat to curved space.
The resulting `curved WDVV equation' is written in terms of a third-rank Codazzi tensor. 
For every flat-space WDVV solution subject to a simple constraint we provide a curved-space solution
on any isotropic space, in terms of the rotationally invariant conformal factor of the metric.
\end{abstract}

\newpage
\pagenumbering{arabic}
\setcounter{page}{1}

\section{WDVV equation in curved space}

A celebrated mathematical structure common to several areas in geometry and mathematical physics
is the (generalized) Witten--Dijkgraaf--Verlinde--Verlinde (WDVV) equation~\cite{W,DVV}
\be\label{fWDVV}
\tF_{kip}\delta^{pq}\tF_{qjm}-\tF_{kjp}\delta^{pq}\tF_{qim}\=0
\qquad\textrm{with}\qquad
\tF_{ijk} \= \pa_i\pa_j\pa_k \tF
\ee
for a real-valued function $\tF(x)$ of $n$ real variables $x=(x^i)=(x^1,x^2,\ldots,x^n)$ 
which can be taken as coordinates of Euclidean space~$\R^n$. 
There is considerable mathematics and physics literature on solutions to this equation.
In this short Letter, we propose a generalization of this equation to any Riemannian manifold~$\cal M$
given by a metric
\be
\diff s^2 \= g_{ik}(x)\,\diff x^i \diff x^k\ .
\ee

In order to motivate the form of the purported `curved WDVV equation', let us 
reformulate the flat-space version by defining $n$ $n{\times}n$ matrices $\htF_i$ 
and a matrix-valued one-form $\htF$ via
\be
\bigl(\htF_i\bigr)^\ell{}_p \=  \delta^{\ell k}\tF_{kip}  \und
\htF \= \diff x^i\,\htF_i \ ,
\ee
respectively~\cite{LST}. In these terms, the WDVV equation~(\ref{fWDVV}) is equivalent to
the flatness or nilpotency condition
\be
\bigl( \diff\pm\htF \bigr)^2 \= 0  \qquad\Leftrightarrow\qquad
\hF\wedge\hF \= 0 \qquad\Leftrightarrow\qquad
\bigl[ \htF_i,\htF_j\bigr] \= 0\ ,
\ee
where both signs are allowed.

This form suggests a natural geometric generalization to curved space, by simply covariantizing
with respect to diffeomorphisms:
\be \label{cWDVV}
\begin{aligned}
\bigl( \diff+\hat\Gamma\pm\hF\bigr)^2 =0 \qquad &\Leftrightarrow\qquad
(\diff+\hat\Gamma)\hF=0 \quad\textrm{and}\quad \hF\wedge\hF+\hat{R} \= 0 \\[4pt]
&\Leftrightarrow\qquad \nabla_{[i}\hF_{j]}\=0 \quad\textrm{and}\quad 
\bigl[ \hF_i,\hF_j\bigr] + \hat{R}_{ij} \= 0\ ,
\end{aligned}
\ee
where
\be
\bigl(\diff+\hat{\Gamma}\bigr)^\ell{}_p \= \diff x^i \bigl(\nabla_i \bigr)^\ell{}_p
\= \diff x^i \bigl( \delta^\ell{}_p \pa_i + \Gamma^\ell{}_{pi} \bigr) \und
\bigl(\hat{R}\bigr)^\ell{}_m \= \diff x^i{\wedge}\diff x^j \bigr( \hat{R}_{ij} \bigr)^\ell{}_m
\= \diff x^i{\wedge}\diff x^j\, R^\ell{}_{mij}
\ee
contain the Levi--Civita connection~$\hat{\Gamma}$, the general coordinate-covariant 
derivative $\nabla_i$ and the Riemann curvature tensor $\hat{R}_{ij}$. 
In components with all indices down, (\ref{cWDVV}) reads
\begin{subequations} \label{cwdvv}
\bea
\nabla_i F_{kjp}- \nabla_j F_{kip} \!\!&=&\!\!0 \\[4pt]
F_{kip}g^{pq}F_{qjm}-F_{kjp}g^{pq}F_{qim} \!\!&=&\!\! -R_{kmij}\ ,
\eea
\end{subequations}
where, as usual and with $g^{\ell k}g_{km}=\delta^\ell{}_m$,
\be
\begin{aligned}
\Gamma^\ell{}_{pi} &\= \sfrac12\,g^{\ell k}\,\bigl( \pa_p g_{ik}+ \pa_i g_{pk} - \pa_k g_{pi} \bigr)\ ,\\[4pt]
R^\ell{}_{pij} &\= \pa_i \Gamma^\ell{}_{pj} - \pa_j \Gamma^\ell{}_{pi}
+ \Gamma^\ell{}_{mi}\,\Gamma^m{}_{pj}-\Gamma^\ell{}_{mj}\,\Gamma^m{}_{pi}\ ,\\[4pt]
\nabla_i F_{jkp} &\= \pa_i F_{jkp} 
-\Gamma^m_{ij} F_{mkp}-\Gamma^m_{ik} F_{jmp}-\Gamma^m_{ip} F_{jkm}\ .
\end{aligned}
\ee
In general, $F_{kip}$ is no longer the third flat derivative of some function (prepotential)~$F$,
but we keep the total symmetry in all three indices,
\be
F_{kip} = F_{ikp} = F_{pik}\ .
\ee
Then, (\ref{cwdvv}a) qualifies $(F_{kip})$ as a so-called third-rank Codazzi tensor~\cite{CodazziT}, 
while (\ref{cwdvv}b) is our curved-space proposal for the WDVV equation.
Note that it is no longer homogeneous in~$F$ but depends on the Riemann tensor of the manifold.
With a more natural index position, it takes the form
\be\label{CWDVV}
F^k{}_{ip}\,F^p{}_{jm}-F^k{}_{jp}\,F^p{}_{im} \= -R^k{}_{mij}\ .
\ee

\setcounter{equation}{0}
\section{${\cal N}{=}\,4$ supersymmetric multi-dimensional mechanics}

Our proposal (\ref{CWDVV}) can in fact be `derived' from the relation of the WDVV equation with 
${\cal N}{=}\,4$ supersymmetric classical (and also quantum) mechanics for 
$n$~nonrelativistic identical particles on a real line. The phase space of this system is given by
$n$ commuting coordinates~$x^i$ and momenta~$p_k$ as well as $4n$ anticommuting variables
\be
\psi^{ia} \und \bar\psi^k{}_b = \bigl(\psi^{kb}\bigr)^\dagger \qquad\text{with}\quad a,b=1,2
\ee
and canonical Poisson brackets between $x$ and $p$ as well as between $\psi$ and $\bar\psi$.
The starting point then is an ansatz for the ${\cal N}{=}\,4$ supercharges,
\be\label{Qf}
Q^a \= p_i \,\psi^{ia} + \im \tF_{ijk}(x)\, \psi^{ib}\, \psi^j{}_b\, \bpsi^{ka} 
\quad\und\quad
\bQ_a \= p_i\, \bpsi^i{}_a   +\im \tF_{ijk}(x)\, \bpsi^i{}_b\, \bpsi^{jb}\, \psi^k{}_a\ ,
\ee
with real structure functions~$\tF_{ink}$ symmetric in the first two indices.
Imposing the supersymmetry algebra
\be\label{N4Poincare}
\left\{ Q^a, Q^b\right\} \=0\= \left\{\bQ_a, \bQ_b \right\} 
\quad\und\quad \left\{ Q^a, \bQ_b\right\} \= \sfrac{\im}{2} \, \delta^a{}_b \,H
\ee
defines the Hamiltonian
\be\label{Hamf}
H \= \delta^{ij} p_i p_j \ -\  
2\bigl( \pa_i \tF_{kjp} + \pa_j \tF_{kip}\bigr) \, {\psi}^{ia}  {\bpsi}^j{}_a\,{\psi}^{kb} {\bpsi}^p{}_b
\ee
and constrains the functions $\tF_{kip}(x)$ to be symmetric in all three indices and
to obey the flat WDVV equations~(\ref{fWDVV}),
as was first demonstrated by Wyllard~\cite{Wyllard} and put into the WDVV context 
by Bellucci, Galajinsky and Latini~\cite{BGL}. In addition, a further condition
\be
\pa_i\tF_{kjp}-\pa_j\tF_{kip} \=0
\ee
is resolved by the existence of the prepotential~$\tF$ in~(\ref{fWDVV}).
Hence, the four-fermion coefficient in brackets in the Hamiltonian~(\ref{Hamf}) equals
$2\pa_i\pa_j\pa_k\pa_p\tF$.

In this context, the curved-space generalization deforms the canonical Poisson brackets to
\be\label{PB}
\big\{ x^i, p_j \big\}= \delta^i_j\ ,\ 
\big\{ \psi^{ia}, \bpsi^j{}_b \big\} = \sfrac{\im}{2}\delta^a{}_b g^{ij}\ ,\
\big\{ p_i, \psi^{aj} \big\} = \Gamma^j_{ik}\psi^{ka}\ ,\
\big\{ p_i, \bpsi^j{}_a \big\} = \Gamma^j_{ik}\bpsi^k{}_a\ ,\
\big\{ p_i, p_j \big\} = -2\im R_{ijkm}{\psi}^{ka}{\bpsi}^m{}_a\ ,
\ee
which unambiguously follows from the first two brackets by employing the Jacobi identities.
We now use exactly the same ansatz (\ref{Qf}) for the supercharges, except that we drop
the $(0)$ superscript on the structure functions,
\be\label{Q}
Q^a \= p_i \,\psi^{ia} + \im F_{ijk}(x)\, \psi^{ib}\, \psi^j{}_b\, \bpsi^{ka} 
\quad\und\quad
\bQ_a \= p_i\, \bpsi^i{}_a   +\im F_{ijk}(x)\, \bpsi^i{}_b\, \bpsi^{jb}\, \psi^k{}_a\ .
\ee
Demanding again the supersymmetry algebra (\ref{N4Poincare}) once more
forces the $F_{kip}$ to be totally symmetric and, using the deformed Poisson brackets~(\ref{PB}),
leads precisely to our curved equations~(\ref{cwdvv}).
The corresponding Hamiltonian acquires the form
\be\label{Ham1}
H\= g^{ij} p_i p_j - 2\bigl( \nabla_i F_{kjp} + \nabla_j F_{kip} + 2 R_{kpij} \bigr)\,
{\psi}^{ia}  {\bpsi}^j{}_a\,{\psi}^{kb} {\bpsi}^p{}_b \ .
\ee

\noindent
{\bf Remark. \ }
If our manifold~$\cal M$ is of constant curvature, i.e.~maximally symmetric,
then a third-rank Codazzi tensor is determined by a single prepotential~\cite{CodazziT},
\be\label{CC1}
\begin{aligned}
F_{ijk} &\= \sfrac13\big(
\nabla_i \nabla_j \nabla_k F+\nabla_j \nabla_i \nabla_k F+\nabla_k \nabla_i \nabla_j F\big)
+ \frac{4 R}{3 n(n{-}1)}\big( g_{jk}\nabla_i F + g_{ik}\nabla_j F + g_{ij}\nabla_k F \big) \\
&\= \nabla_i \nabla_j \nabla_k F
+ \frac{R}{n(n{-}1)} \big( 2 g_{jk} \partial_i F + g_{ik} \partial_j F + g_{ij}\partial_k F \big)
\qquad\textrm{with}\quad R = R^\ell{}_{j\ell m}\,g^{jm}\=\textrm{const.} \ .
\end{aligned}
\ee
Having solved (\ref{cwdvv}a) in terms of~$F(x)$, the curved WDVV equation~(\ref{cwdvv}b) yields
a rather complicated equation for this prepotential, which we have not investigated further.
Instead, we have found large classes of solutions on isotropic (not necessarily
homogeneous) spaces, which we shall display in the remainder of this Letter.

\newpage

\setcounter{equation}{0}
\section{Particular solutions of curved WDVV equation}

\subsection {Potential metric}
Motivated by the results of \cite{AP1, AP2} on ${\cal N}{=}\,4$ $n$-dimensional supersymmetric mechanics, 
we consider metrics given by a potential function,
\be\label{RK1}
g_{ij} \= \pa_i \pa_j F(x)\ .
\ee 
For a manifold admitting such a metric, one finds that
\be\label{RKGammaRiem}
\Gamma_{ijk} \= \sfrac12\pa_i\pa_j\pa_k F \quad\und\quad 
R_{ijkm} \= g^{pq} \big( \Gamma_{imp}\Gamma_{jkq} - \Gamma_{ikp}\Gamma_{jmq}  \big) \ .
\ee
It is then rather easy to see that the choice
\be\label{RKF}
F_{ijk} \= \pm \Gamma_{ijk}
\ee
solves both curved WDVV equations (\ref{cwdvv}) for any choice of the potential~$F$.

\subsection{Isotropic}
Suppose that the manifold $\cal M$ is isotropic, 
i.e.~it has $\sfrac12n(n{-}1)$ Killing vectors and admits an SO$(n)$-invariant metric,
\be
\diff s^2 \= g_{ik}(r)\,\diff x^i \diff x^k \qquad\textrm{for}\quad r^2 = \delta_{ij} x^i x^j\ .
\ee
Such a space is a cone over $S^{n-1}$ and is conformally flat, 
so we can find coordinates in which the metric takes the form
\be
\diff s^2 \= \diff r^2 + h(r)^2 \diff\Omega_{n-1}^2 
\qquad\textrm{or}\qquad
\diff s^2 \= f(r)^{-2} \delta_{ik} \diff x^i \diff x^k\ ,
\ee
where $\diff\Omega_{n-1}^2$ is the metric on the round $(n{-}1)$-sphere.

Let us make the following ansatz, inspired by~\cite{GLP},
\be\label{Fansatz}
F_{ijk} \= a(r)\,x^i x^j x^k + b(r)\bigl(\delta_{ij}x^k+\delta_{jk}x^i+\delta_{ki}x^j\bigr) 
+ c\,f(r)^{-2}\tF_{ijk}(x) 
\ee
including an arbitrary flat-space solution $\tF$ 
(the $f^{-2}$ prefactor arises from pulling down the first index of~$F$).
The two functions $a(r)$ and $b(r)$ are to be determined depending on the constant~$c$.
One may check that the linear equation~(\ref{cwdvv}a) is satisfied if 
\be\label{mWDVV1}
r(rf'-f)\,a + 4f'b + f\,b'\= 0 \quad\und\quad x^i\tF_{ijk}\ \= \delta_{jk}\ ,
\ee
where we fixed the scale of $\tF$. Here, prime means differentiation with respect to~$r$.

The curved WDVV equation (\ref{cwdvv}b) further imposes the quadratic conditions
\be \label{mWDVV2}
f^2 b \left( r^2 a +b\right) + c\,a \= -\frac1{r f^3}\Bigl(\frac{f'}{r}\Bigr)' \quad\und\quad
r^2 f^2 b^2 + 2\,c\,b \= \frac{r^2 f'}{f^4} \Bigl(\frac{f}{r^2}\Bigr)'\ .
\ee
Interestingly, these equations already imply the condition~(\ref{mWDVV1}), for any value of~$c$,
meaning that (\ref{cwdvv}a) follows from (\ref{cwdvv}b) for our ansatz~(\ref{Fansatz}).
The equations \p{mWDVV2} may be easily solved as
\be \label{solmWDDV}
\begin{aligned}
a &\= \frac{2 c f \sqrt{c^2 f^2 - 2r f f' + r^2(f')^2} \pm \bigl( 2c^2f^2 - 3r f f' + r^2(f')^2 + r^2 f f''\bigr)}
{r^4 f^3 \sqrt{c^2 f^2 - 2r f f' + r^2(f')^2}} \ ,\\
b &\= -\frac{c f \pm \sqrt{c^2 f^2 - 2r f f' + r^2(f')^2}}{r^2 f^3} \ .
\end{aligned}
\ee
For $c{=}1$, it simplifies to
\be
a \= \frac{2 f (\sfrac{f}{r})' \mp r \bigl(f (\sfrac{f}{r})'\bigr)'}{r^4 f^3 (\sfrac{f}{r})'}
\quad\und\quad
b \= -\frac{f \mp r^2(\sfrac{f}{r})'}{r^2 f^3} \ .
\ee
Thus, for isotropic metrics any flat-space WDVV solution~$\tF$ 
may be lifted to a solution of the curved WDVV equation via~\p{Fansatz}.

\setcounter{equation}{0}
\section{Conclusion}
We have employed the relation between $n$-dimensional ${\cal N}{=}\,4$ supersymmetric mechanics and the WDVV equation
to generalize the latter to curved spaces, i.e.~to arbitrary Riemannian manifolds.
In this `curved WDVV equation', the third derivative of the prepotential, $\tF_{ijk}=\pa_i\pa_j\pa_k\tF$ 
is replaced by the third-rank Codazzi tensor $F_{ijk}$, while the WDVV equation itself acquires a non-trivial right hand side
given by the Riemann curvature tensor. 
We have found solutions of the curved WDVV equation for metrics with a potential and on arbitrary isotropic spaces. 
The latter solution is built on an arbitrary solution $\tF$ of the {\it flat\/} WDVV equation subject to $x^i\tF_{ijk}\sim\delta_{jk}$. 
Thus, any such flat solution can be lifted to a curved solution on an isotropic space.

Here, we have worked with a restricted form of ${\cal N}{=}\,4$ supersymmetric mechanics, with vanishing potential $W$.
So the obvious reverse application to multi-dimensional supersymmetric mechanics will extend the supercharges by the
potential terms (as it was done in \cite{Wyllard}, \cite{GLP}, \cite{KL1} for the flat case) and to find admissible potentials. 
This task will be considered elsewhere. 

\section*{Acknowledgements}
We are grateful to M.~Feigin and A.~Veselov for stimulating discussions.
A.S.\ also thanks S.~Kuzenko, A.~Sagnotti and D.~Sorokin for 
valuable comments and discussions during the Ginzburg conference in Moscow.
This work was partially supported by the Heisenberg-Landau program.
The work of N.K.\ and S.K.\ was partially supported by RSCF grant 14-11-00598,
the one of A.S.\ by RFBR grant 15-02-06670. 
The work of A.N. was partially supported by the Armenian State Committee of Science Grant 15T-1C367.
This article is based upon work from COST Action MP1405 QSPACE, supported by COST (European Cooperation in Science and Technology).



{\small
\begin{thebibliography}{99}

\bibitem{W} E.~Witten, \\
{\it On the structure of the topological phase of two-dimensional gravity},\\
Nucl. Phys. B {\bf 340} (1990) 281.

\bibitem{DVV}
R.~Dijkgraaf, H.~Verlinde, E.~Verlinde, \\
{\it Topological strings in $d<1$},\\
Nucl. Phys. B {\bf 352} (1991) 59.

\bibitem{LST} O.~Lechtenfeld, K.~Schwerdtfeger, J.~Th\"urigen, \\
{\it ${\cal N}{=}\,4$ multi-particle mechanics, WDVV equation and roots},\\
SIGMA {\bf 7} (2011) 023, {\tt arXiv:1011.2207 [hep-th]}.

\bibitem{CodazziT} H.L.~Liu, U.~Simon, C.P.~Wang, \\
{\it Higher order Codazzi tensors on conformally flat spaces},  \\
Contributions to Algebra and Geometry {\bf 39} (1998) 329.

\bibitem{Wyllard} N.~Wyllard,\\
{\it (Super)conformal many-body quantum mechanics with extended supersymmetry},\\
J. Math. Phys. {\bf 41} (2000) 2826, {\tt arXiv:hep-th/9910160}.

\bibitem{BGL} S.~Bellucci, A.V.~Galajinsky, E. Latini,\\
{\it  New insight into the Witten--Dijkgraaf--Verlinde--Verlinde equation},\\
Phys. Rev. D {\bf 71} (2005) 044023, {\tt arXiv:hep-th/0411232}.

\bibitem{AP1} E.~Donets, A.~Pashnev, J.~Juan Rosales, M.~Tsulaia, \\
{\it N=4 supersymmetric multidimensional quantum mechanics, partial SUSY breaking\\
and superconformal quantum mechanics},\\
Phys. Rev. D {\bf 61} (2000) 043512, {\tt arXiv:hep-th/9907224}.
  
\bibitem{AP2} E.E.~Donets, A.~Pashnev, V.O.~Rivelles, D.~Sorokin, M.~Tsulaia,\\
{\it N=4 Superconformal mechanics and the potential structure of AdS spaces}, \\
Phys. Lett. B {\bf 484} (2000) 337, {\tt arXiv:hep-th/0004019}.

\bibitem{GLP} A.~Galajinsky, O.~Lechtenfeld, K.~Polovnikov, \\ 
{\it N=4 superconformal Calogero models}, \\
JHEP {\bf 0711} (2007) 008, {\tt arXiv:0708.1075[hep-th]}.

\bibitem{KL1} S.~Krivonos, O.~Lechtenfeld, \\
{\it Many-particle mechanics with $D(2,1;\alpha)$ superconformal symmetry}, \\
JHEP {\bf 1102} (20112) 042, {\tt arXiv:1012.4639[hep-th]}.

\end{thebibliography}
}
\end{document}